\documentclass{elsart}
\usepackage{amsmath}
\usepackage{amssymb}
\usepackage{epsfig}
\usepackage{psfrag}

\newcommand{\mbf}{\mathbf}

\begin{document}
\begin{frontmatter}

\title{\hfill{\tiny FZJ--IKP(TH)--2005--32} \\[1.8em]
Insights on scalar mesons from their radiative decays}
 
\author{Yu. Kalashnikova$^1$, A. Kudryavtsev$^1$, A. V. Nefediev$^1$,}
\author{ J. Haidenbauer$^2$, C. Hanhart$^2$}

{\small $^1$Institute of Theoretical and Experimental Physics,
117218,}\\
{\small B.\-Cheremushkinskaya 25, Moscow, Russia} \\
{\small $^2$Institut f\"{u}r Kernphysik, Forschungszentrum J\"{u}lich GmbH,}\\ 
{\small D--52425 J\"{u}lich, Germany} 

\maketitle

\begin{abstract}
\noindent We estimate the rates for radiative transitions of the
lightest scalar mesons $f_0(980)$ and $a_0(980)$
to the vector mesons $\rho$ and $\omega$. 
We argue that measurements of the radiative decays of those
scalar mesons can provide important new information on their structure.
\end{abstract}

\end{frontmatter}

\section{Introduction}

Although studied since many decades, the lightest scalar mesons and,
especially the $f_0(980)$ and $a_0(980)$, are still
subject of debate regarding their fundamental structure. 
For example, these two mesons could be viewed
as natural candidates for the genuine $1^3P_0$ $q \bar q$ states
predicted by the standard quark models \cite{qq}.
However, due to the proximity of these states to the $K \bar K$
threshold, a significant if not dominant $qq \bar q \bar q$ configuration is 
expected from a phenomenological point of view. Thus, it was argued (see, e.g.,
Ref. \cite{CT}) that the genuine $q \bar q$ $^3P_0$ nonet could be somewhere 
near $1.5$ GeV, while the states around $1$ GeV are due to a strong $S$-wave 
attraction between the two quarks and two antiquarks. 

In such a scenario the $f_0(980)$ and $a_0(980)$ mesons could 
be realized either in form of 
compact $qq \bar q \bar q$ states \cite{Jaffe,4q} or in form 
of loosely bound $K \bar K$ states. To complicate things further, 
within the latter picture there are even two possibilites 
regarding the nature of those scalar resonances, which are
connected with the question whether there are sufficiently strong
$t$--channel forces so that $K\bar K$ molecules are formed,
as advocated in Refs. \cite{WeiIs,Jue,Oset,Markushin}, or whether the
meson--meson interaction is dominated by s--channel states. In the
former case the $f_0(980)$ and $a_0(980)$ would be purely composite
particles, whereas in the latter case they would contain both 
elementary states and composite-particle components. 
For a much more thorough discussion on that issue and 
an overview of the extensive literature we refer the reader to 
the reviews in Refs. \cite{Buggrep,Klempt,Amsl,Oll,Bev,Ani}. 

Over the years
many experiments have been proposed in order to distinguish among those
scenarios, however, so far the smoking gun experiment has not been identified yet.
For example, about a decade ago it was believed that data on the decays
$J/\Psi \to \phi\pi\pi/\phi KK$ would allow to resolve the puzzle of the
scalar mesons \cite{Penning}. 
Even earlier Achasov and Ivanchenko 
had argued that measurements of the radiative decays of the
$\phi (1020)$ to scalar mesons would provide decisive information on the
structure of these long--debated objects \cite{AI}. The authors of that work
demonstrated that the spectrum of, {\em e.g.}, $\pi^0\pi^0$ in the reaction
$\phi \to \gamma S\to\gamma \pi^0\pi^0$ would look drastically different in
the presence or absence of a significant $K\bar K$ contribution, due to the
proximity of both the mass of the scalar meson $S$ and that of the $\phi$ to
the $K\bar K$ threshold. And indeed, the data \cite{SND,CMD,KLOE}
unambiguously show a prominent $K \bar K$ contribution.
Based on large--$N_C$ considerations, this was interpreted then as a proof
for a compact four--quark nature of the scalar mesons \cite{Achasov}.
However, it is not clear {\em a priori} how quantitative the large--$N_C$
counting rules are in the scalar sector --- for example, the large--$N_C$
analysis of the unitarised chiral perturbation theory amplitudes leads to
large uncertainties for the $a_0/f_0(980)$ states \cite{Pelaez}. After
all, it is well known that, for the scalars, due to the presence of the
nearby strong $S$-wave $K \bar K$ channel, unitarity corrections are
large, as seen from the corresponding Flatt{\' e} distributions
\cite{Flatte,Bugg}. Thus, the large--$N_C$ picture might be obliterated.
In particular, the admixture of a $K \bar K$ component in the scalar wave
function should be large, as discussed in
Refs.~\cite{WeiIs,Jue,Tornqvist,W}.

For obvious reasons, a dominant role of the $K \bar K$ component in the 
$\phi \to \gamma S$ decays is naturally expected in the scenario where the 
scalar mesons are $K\bar K$ molecules. Thus, it might be not too surprising
that explicit calculations utilizing such molecular models 
\cite{Oller,Marco,Markushin} were
able to describe the spectrum of the radiative $\phi$ decays.
Indeed, it was shown \cite{radmol} recently on rather general grounds that, 
contrary to earlier claims \cite{CIK,AGS}, the available experimental information 
is completely consistent with the molecular structure of the scalar mesons. 
We thus conclude that the radiative $\phi$ decays measure the molecular
component of the scalar mesons. However, other observables are to be found
that allow one to understand how much compact structure there is in 
addition.

There is a drawback of the radiative $\phi$ decays: beyond the
prominence of the kaon loops, no further model--independent quantitative
conclusion on the scalar mesons is possible because of the limited phase space
available for these decays. In addition, gauge invariance forces the spectrum,
for large pseudoscalar invariant masses, to behave as $\omega^3$, where
$\omega$ denotes the photon energy. With the photon energy being just 
around $40$ MeV, only a small fraction of the spectral functions of the scalar
mesons is visible in these reactions. As discussed in detail in Ref.~\cite{Pennington}, 
this causes uncertainties in the attempts to define the
coupling constants and pole positions of the scalars.

In view of the difficulties outlined above, with the present paper,
we would like to draw attention to another class of radiative
decays --- namely, to the radiative decays of the scalar mesons themselves. 
In particular, we want to provide evidence for
the following properties of the reactions $S\to \gamma V$, where $S$ denotes the
scalar mesons $a_0$ or $f_0$ and $V$ stands for the vector meson $\rho$ or $\omega$:
\begin{enumerate} 
\item both quark loops and meson loops can be of equal importance;
\item there is significant phase space available for the final state;
\item since there is a sensitivity to the nonstrange contribution of the
wave functions, a combined analysis of the $\phi$ radiative decays, as well as 
those of the scalars should help to map out the underlying 
quark structure of the latter.
\end{enumerate} 

Among those the first point is specifically interesting. It implies that,
if the scalar mesons were predominantly $q\bar q$ or $qq\bar q\bar q$ states,
then quark loops as well as meson loops should yield sizable contributions to 
the decay amplitude and, consequently, they should have a significantly larger 
decay rate to vector mesons as compared to $K\bar K$ molecules,
where only meson ($K\bar K$) loops are present. 
 
In this context we also consider the decay $S\to \gamma \gamma$ and show 
that, in this case, there is again a striking difference in the reaction mechanism
in the sense that now the quark loops dominate while the meson loops
are suppressed. As a consequence, there is a certain pattern or 
hierarchy in the studied radiative decay reactions involving scalar mesons 
($\phi\to \gamma S$, $S\to \gamma V$, $S\to \gamma \gamma$) 
with characteristic differences for a compact ($q\bar q$ or $qq\bar q\bar q$)
or molecular structure of those scalars. 
This suggests that a combined analysis of such decays, within a 
specific scenario of the scalar mesons, is actually a much more conclusive
method to discriminate between these scenarios than just considering a 
single decay mode, like $\phi\to \gamma S$, as it happened in the past. 

The paper is structured in the following way: 
In Sect. 2, we provide general expressions for the vertex function involving a photon, a
scalar ($S$), and a vector ($V$) meson and for the total width of the
transition $S \to \gamma V$.
In Sect. 3, we consider different decay mechanisms 
for the scalar mesons, {\em i.e.}, $q\bar q$ and $qq\bar q\bar q$ quark 
loops and meson loops, and we evaluate the decay width within the 
corresponding transition mechanisms. 
Sect. 4 is devoted to the decay of the scalar
mesons to the $\gamma\gamma$ channel. 
Our results are analysed and discussed thoroughly in Sect. 5. 
The paper ends with a brief summary.  

\section{Some generalities}

In order to write down the effective vertex for the $SV\gamma$ coupling, one is 
to respect gauge invariance for the photon. This is most easily 
implemented by using the field strength tensor for the latter. Therefore,  
the most general structure of the $SV\gamma$ vertex 
is\footnote{Here the standard normalisation of the invariant
amplitude is used, like, {\em e.g.}, in Ref~\cite{CIK}.}
\begin{equation}
iW=M(p^2,q^2)[(k \cdot \epsilon^{V*})(q \cdot \epsilon^{\gamma *})-
(\epsilon^{V*} \cdot \epsilon^{\gamma *})(k \cdot q)],
\label{vertex} 
\end{equation}
where $\epsilon^{V}_\mu$ and $\epsilon^{\gamma}_\mu$ are the polarisation vectors of the
vector meson and photon, $q_\mu$ and $k_\mu$ are their four--momenta,
respectively, and $p_\mu$ is the scalar four--momentum.  For the $\phi$
radiative decays, the decay amplitude exhibits a strong $p^2$--dependence, due
to the proximity of the $K\bar K$ threshold to both the $\phi$ mass as well as
to the nominal mass of the scalar meson. However, for these decays, we have
$q^2=m_\phi^2$, and thus it is not possible to investigate the 
$q^2$--dependence.  On the other hand, for the decays $S\to\gamma V$ and the case of
kaon loop contributions, $M(p^2,q^2)$ shows a significant dependence on both
$p^2$ and $q^2$, due to the proximity of the $K \bar K$ threshold to the mass
of the $a_0/f_0$ mesons and due to the finite width of the vector mesons,
especially of the $\rho$--meson.

For stable scalar and vector mesons one could directly deduce the expression for 
the total width of the transition $S \to \gamma V$ from Eq.~(\ref{vertex}),
which would be given by 
\begin{equation}
\Gamma(m_S^2) =
\frac{m_S^3}{32\pi}|M(m_S^2,m_V^2)|^2\left(1-\left(\frac{m_V}{m_S}\right)^2\right)^3,
\label{observ}
\end{equation}
with $m_V$ and $m_S$ being the nominal masses of the vector and scalar mesons.
For the calculation of observables, in addition to the matrix element $M$, two more 
ingredients are relevant --- namely, the propagator of the scalar
meson $D_S(p^2)$ and that of the vector meson, $D_V(q^2)$. The latter modifies the 
invariant mass spectrum of the final state, cf. the detailed discussion in the
appendix. 

The finite width of the scalar mesons makes one study the decay rates 
as a function
of the invariant mass of the decaying system.
Consequently, in the total decay width, $D_S(m^2_S)$ appears as a weight
factor. For this distribution, one would need to use parametrizations given 
in the literature. Note that using such parametrizations one 
could run into complications connected to possible interference effects
between the $f_0(980)$ and the broad $I=0$ $\pi \pi$ component usually
referred to as ``$\sigma$'' \cite{Buggp}.
In what follows we do not consider this possibility and give estimates 
for stable vectors and scalars.

As mentioned before, it is the transition matrix element $M$ which is the quantity 
of interest, and we
investigate it now in more detail for various scenarios.

\section{The transition matrix element $M$}

In this section, we discuss the properties of $M$ in various models for 
the scalar mesons and for various mechanisms of the radiative decay. 

\subsection{Contribution of quark loops}

\begin{figure}[t]
\begin{center}
\epsfig{file=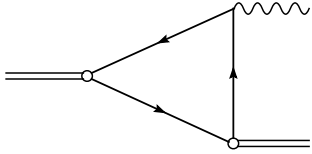, height=2cm}
\caption{\it Quark loop contribution to the radiative decay.}
\label{qqlo}
\end{center}
\end{figure}

The simplest assignment for the $a_0(980)/f_0(980)$ mesons is the bound 
$q\bar{q}$ $^3P_0$ state \cite{qq}. Correspondingly, the radiative decay proceeds 
via a quark loop, as displayed at Fig.~\ref{qqlo}. If confinement is modeled
by a quark--antiquark interaction, then the ingredients needed to
calculate the transition matrix element are: i) the meson--quark--antiquark
vertices, ii) the dressed propagators of quarks, and iii) the dressed
photon--quark--quark vertex. Only if the underlying quark model provides
these ingredients in a selfconsistent way, then the electromagnetic 
transition vertex is compatible with e.m. gauge invariance
and the $SV\gamma$ transition amplitude takes the form of Eq.~(\ref{vertex}).
Reliable calculations of the quark loop contributions can be done in
the framework of the nonrelativistic quark model. The radiative transition 
$^3S_1(q \bar q)\leftrightarrow {}^3P_0(q \bar q)$  
is an $E1$ transition, and the current 
in the rest frame of the initial meson $i$, in the lowest approximation, is 
\begin{equation}
{\mbf j}_{i\to f}=\left\langle f\left|e_q\frac{{\mbf p}_{q}}{m_q}
\right|i\right\rangle+(q\leftrightarrow\bar q).
\label{current}
\end{equation}

The expression for the matrix element $M$, extracted from Eq.~(\ref{current}), reads
\begin{equation}
M=\frac{2}{3}e\langle Q\rangle r_{if},
\label{nrquark}
\end{equation}
where 
\begin{equation}
Q=\frac12(Q_q-Q_{\bar q})
\end{equation}
is the quark charge operator, and
the radial part of the dipole matrix element between the initial and final states reads
\begin{equation}
r_{if}=\int r^2dr R_f^\dagger(r)r R_i(r),
\label{rif}
\end{equation}
with $R_{i,f}(r)$ being the radial wave functions for the initial and the final states, 
respectively.

Generally, the decay rate for $E1$ transitions between the 
$^3S_1$ and $^3P_J$ states is given by (see, {\em e.g.}, Ref.~\cite{KR})
\begin{equation}
\Gamma=\frac{4(2J+1)}{27}\alpha \langle Q\rangle^2 \omega^3 r^2_{if},
\label{sp}
\end{equation}
for the $^3S_1 \to ^3P_J \gamma$ decays, and by
\begin{equation}
\Gamma=\frac{4}{9}\alpha \langle Q\rangle^2 \omega^3 r^2_{if}
\label{ps}
\end{equation}
for the $^3P_J \to ^3S_1 \gamma$ decays. Here $\omega$ stands for the photon energy and 
the charge factor is readily calculated for a given flavour of
the initial and final states ($n$ denotes $u$ and/or $d$ quark):
\begin{equation}
\langle Q\rangle^2=
\left\{
\begin{array}{ccll}
\frac{1}{36},&\rm for&n{\bar n} \to n{\bar n}&\rm with~the~same~isospin,\\
\frac{1}{4},&\rm for&n{\bar n} \to n{\bar n}&\rm with~different~isospins,\\
\frac{1}{9},&\rm for&s{\bar s} \to s{\bar s}.&
\end{array}
\right.
\label{iso}
\end{equation}

One might question the applicability of the nonrelativistic or the naively
relativised quark model to the mentioned decays. Nevertheless, experimental data
can be used to estimate the needed matrix element. We may use the known
radiative decay rate of the {\it bona fide} quarkonium $f_1(1285)$ \cite{PDG}, as a
genuine $^3P_1$ $q \bar q$ state made of light quarks:
\begin{equation}
\Gamma(f_1(1285)) \to \gamma \rho)=1320 \pm 312 {\rm keV}.
\label{f1}
\end{equation}
As shown in Ref.~\cite{CDK}, nonrelativistic quark models with standard parameters
yield results for the decay $f_1(1285) \to \gamma \rho$ that are in good
agreement with the data. To relate this matrix element to the ones of
interest we assume $SU(6)$ symmetry for the wave functions which is
expected to provide a reasonable order--of--magnitude estimate for the rates.
In this case the values of the matrix elements $r_{if}$ are to be equal for 
all members of the $P$--multiplet. 

Then one gets from Eqs.~(\ref{sp})-(\ref{f1}):
\begin{equation}
\begin{array}{l}
\Gamma(a_0 \to \gamma \omega) = \Gamma(f_0(n \bar n) \to \gamma \rho)=125~{\rm keV},\\
\Gamma(a_0 \to \gamma \rho) = \Gamma(f_0(n \bar n) \to \gamma \omega)=14~{\rm keV},\\
\Gamma(f_0(s \bar s) \to \gamma \omega) = 31~{\rm keV} \times\sin^2\theta,\\
\Gamma(f_0(s \bar s) \to \gamma \rho)=0,\\
\end{array}
\label{V}
\end{equation}
where $\sin\theta$ measures the (small) $\phi-\omega$ mixing. 

While the expressions (\ref{sp}) and (\ref{ps}) take apparently 
nonrelativistic form, relativistic corrections are actually 
included in these dipole formulae, provided the masses and the wave functions of
the initial and final mesonic states are taken to be solutions of a
quark--model Hamiltonian with relativistic corrections taken into account (see 
Refs.~\cite{McClary,Orsay} for a detailed discussion).
With relativistic
corrections to the wavefunctions taken into account the values of $r_ij$
for $^3P_0$ and $3^P_1$ states are not equal to each other anymore. So the
estimates (\ref{V}) are to be considered as order-of-magnitude ones.

Similarly we obtain for $\phi$ decay: 
\begin{equation}
\begin{array}{l}
\Gamma(\phi \to \gamma a_0) =0.37~{\rm keV} \times\sin^2\theta,\\
\Gamma(\phi \to \gamma f_0(s \bar s)) =0.18~{\rm keV},\\
\Gamma(\phi \to \gamma f_0(n \bar n)) =0.04~{\rm keV}\times\sin^2\theta.\\
\end{array}
\label{phi}
\end{equation}
In this context let us mention that 
the pure $s \bar s$ assignment for $f_0$ seems implausible as 
it implies 
an OZI suppression of the $\pi\pi$ mode in the $\phi$ radiative decay, 
so that some mixing with an $n \bar n$ isoscalar state is needed to 
reproduce the branching fraction of the $f_0(980)$ to $\pi \pi$.
 
We would like to point out that the predictions for the decay width ratios 
of rates,
\begin{equation}
\begin{array}{c}
\Gamma[(f_0(n \bar n) \to \gamma \rho_0):(a_0 \to \gamma \omega):(f_0(n
\bar n) \to \gamma \omega):(a_0 \to \gamma \rho_0)]=\\
9:9:1:1,
\end{array}
\label{quarkratio}
\end{equation}
which are readily deduced from Eq.~(\ref{V}), are based only on the 
isospin relations (\ref{iso}) and are, therefore, robust. 

\subsection{Contributions of annihilation graphs for $qq \bar q \bar q$ scalars}

\begin{figure}[t]
\begin{center}
\epsfig{file=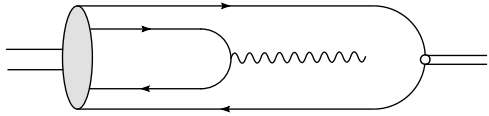, height=1.5cm}
\caption{\it Annihilation contributions to the radiative decay
of $qq\bar q\bar q$ scalars.}
\label{4qlo}
\end{center}
\end{figure}
  
In the diquark--antidiquark model \cite{Jaffe,4q}
both $f_0$ and $a_0$ can be identified with $sn\bar{s}\bar{n}$ states
belonging to a cryptoexotic $\overline{3}\otimes 3$ flavour nonet,
\begin{align}
f_0=\frac{1}{2}([su][\bar s \bar u]+[sd][\bar s \bar d]),\nonumber \\[-5mm]
\label{4q}\\
a_0(I_3=0)=\frac{1}{2}([su][\bar s \bar u]-[sd][\bar s \bar d]).\nonumber
\end{align}

The radiative decays of the $qq \bar q \bar q$ states proceed via annihilation 
of a $q \bar q$ pair, as shown at Fig.~\ref{4qlo}. Thus, in the transition 
$S \to \gamma \rho/\omega$, the $s \bar s$ pair annihilates, so that one has
\begin{equation}
\frac{\Gamma (f_0 \to \gamma \rho)}{\Gamma (f_0 \to \gamma \omega)}=0,
\end{equation}
while the $\phi-\omega$ mixing could generate small nonzero values of the ratio
\begin{equation}
\frac{\Gamma (a_0 \to \gamma \omega)}{\Gamma (a_0 \to \gamma \rho)} \sim \sin^2\theta.
\end{equation}
On the contrary, the decay $\phi \to \gamma S$ with the four--quark scalars (\ref{4q}) 
proceeds via creation of a $n \bar n$ pair (if
one neglects the $\phi-\omega$ mixing), yielding
\begin{equation}
\frac{\Gamma(\phi \to \gamma a_0)}{\Gamma(\phi \to \gamma 
f_0)}=9.
\label{4qphi}
\end{equation}
Note that the experimental value for this ratio is around 1/6 \cite{CDK}. 
    
The assumption (\ref{4q}) is compatible with the $a_0/f_0$ mass degeneracy. 
However, with the $sn\bar{s}\bar{n}$ assignment for $f_0$, a superallowed 
decay to $\pi\pi$ is impossible, so that one is forced to assume a mixing of
the isoscalar $sn\bar{s}\bar{n}$ state with a $\sigma$-like $nn\bar{n}\bar{n}$
state (see \cite{4q}). 
Note that there is no such problem for the superallowed decay $a_0 \to \pi \eta$, since
the $\eta$ contains an admixture of the strange quark pair.

There are no theoretical estimates of the absolute values for the $qq \bar q \bar q$ 
radiative decay rates. Moreover, since no single four--quark state is 
unambiguously identified, there is also no experimental anchor at our
disposal, similar to Eq.~(\ref{f1}), which could allow one to predict absolute 
values of these rates.

\subsection{Contribution of meson loops}

\begin{figure}[t]
\begin{center}
\begin{tabular}{ccc}
\epsfig{file=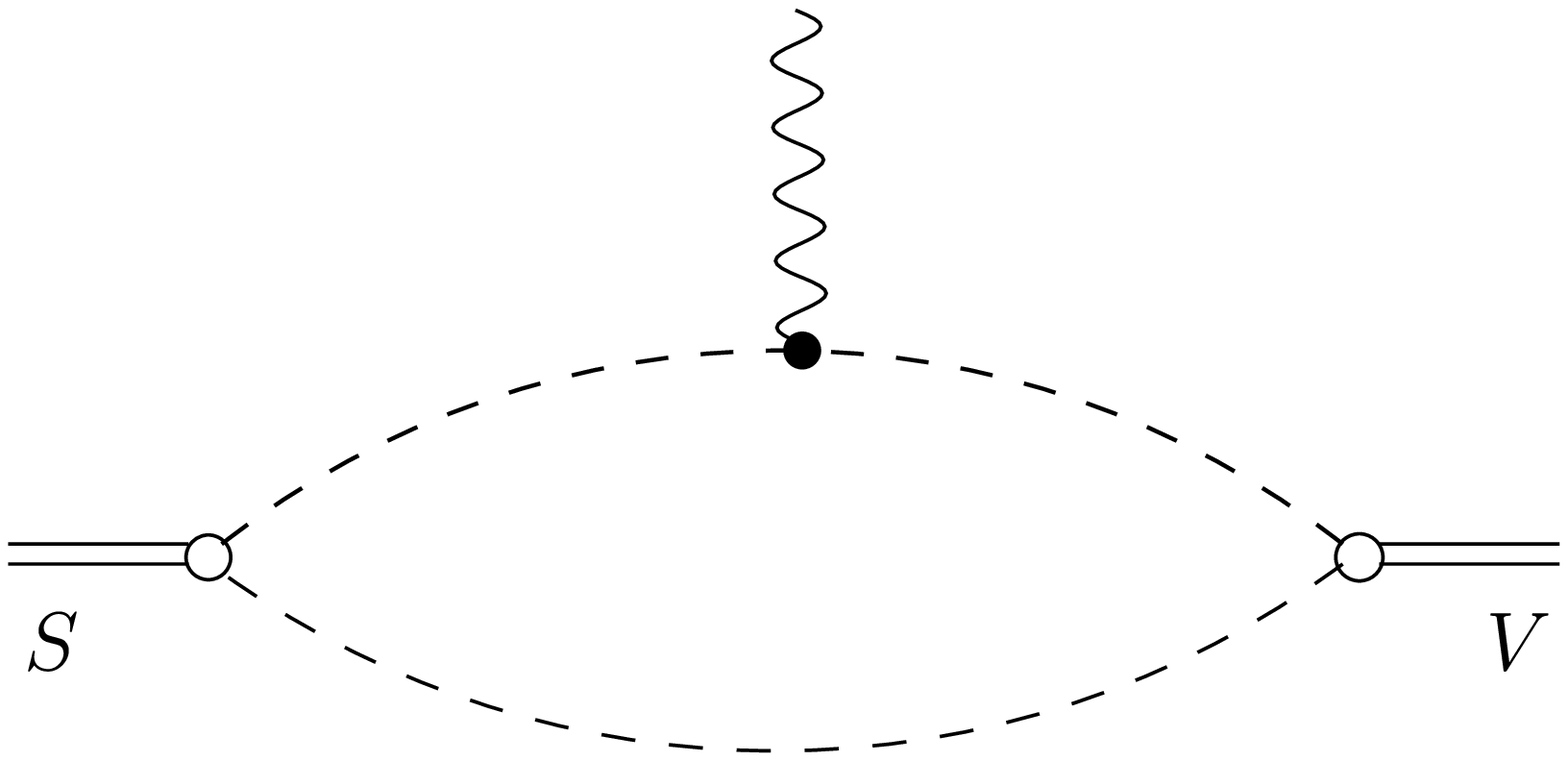,width=4cm}&
\raisebox{-8mm}{$\epsfig{file=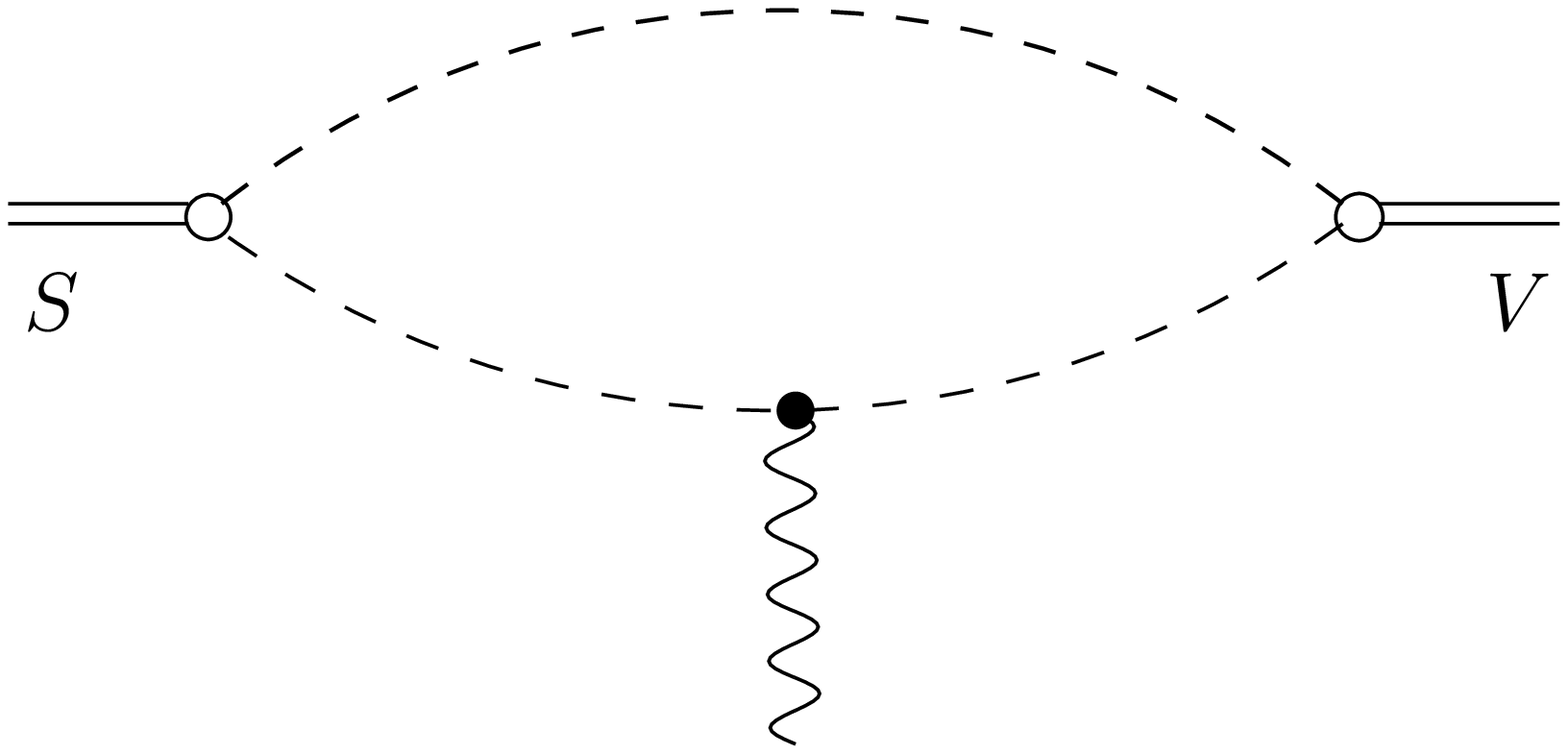,width=4cm}$}&
\epsfig{file=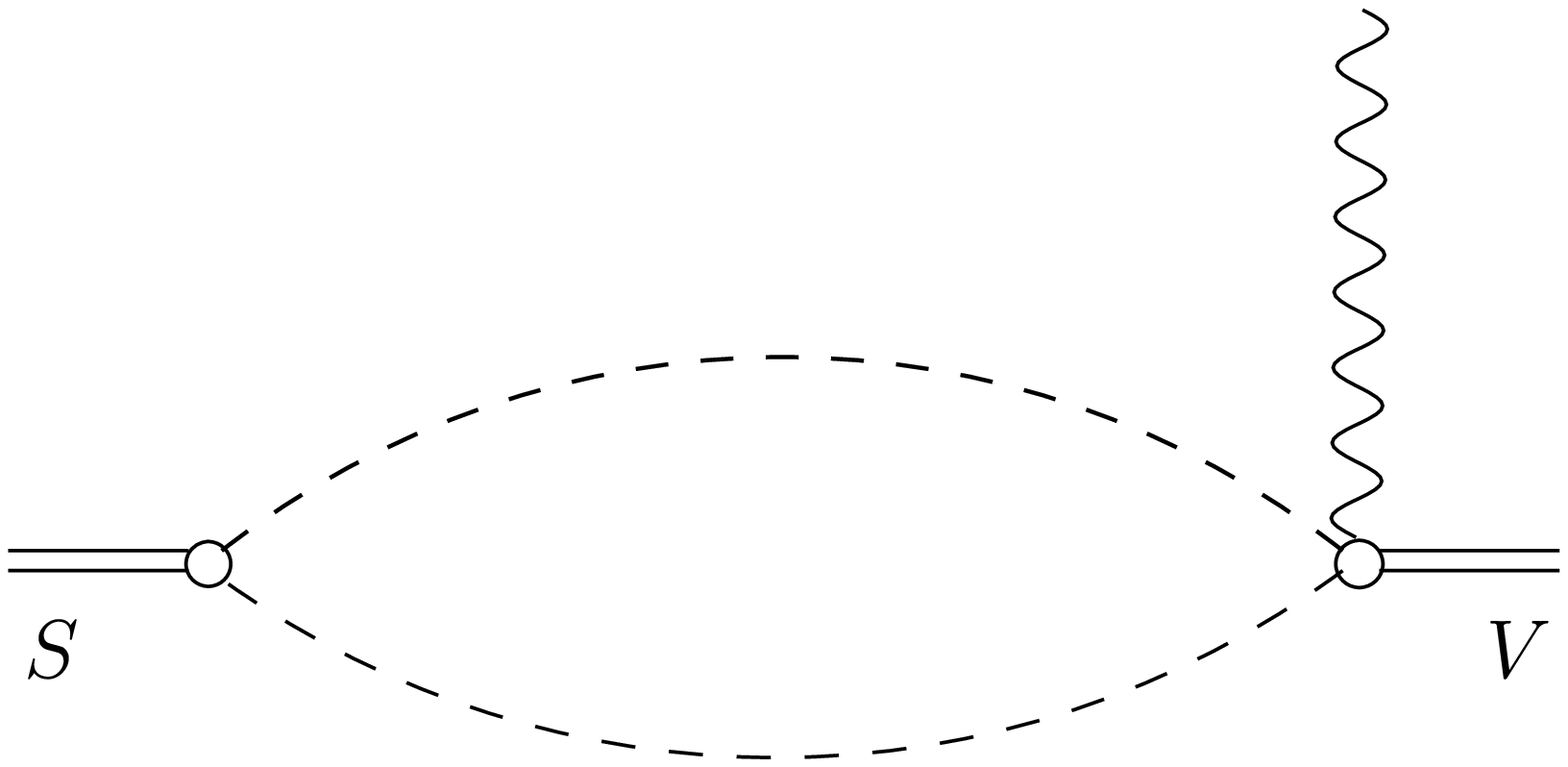,width=4cm}\\
a)&b)&c)
\end{tabular}
\caption{\it Meson loop contributions to the radiative decay of scalar mesons. 
Diagram c) is required to provide the overall gauge invariance of the amplitude.}
\label{melo}
\end{center}
\end{figure}

\begin{figure}[t]
\begin{center}
\centerline{\epsfig{file=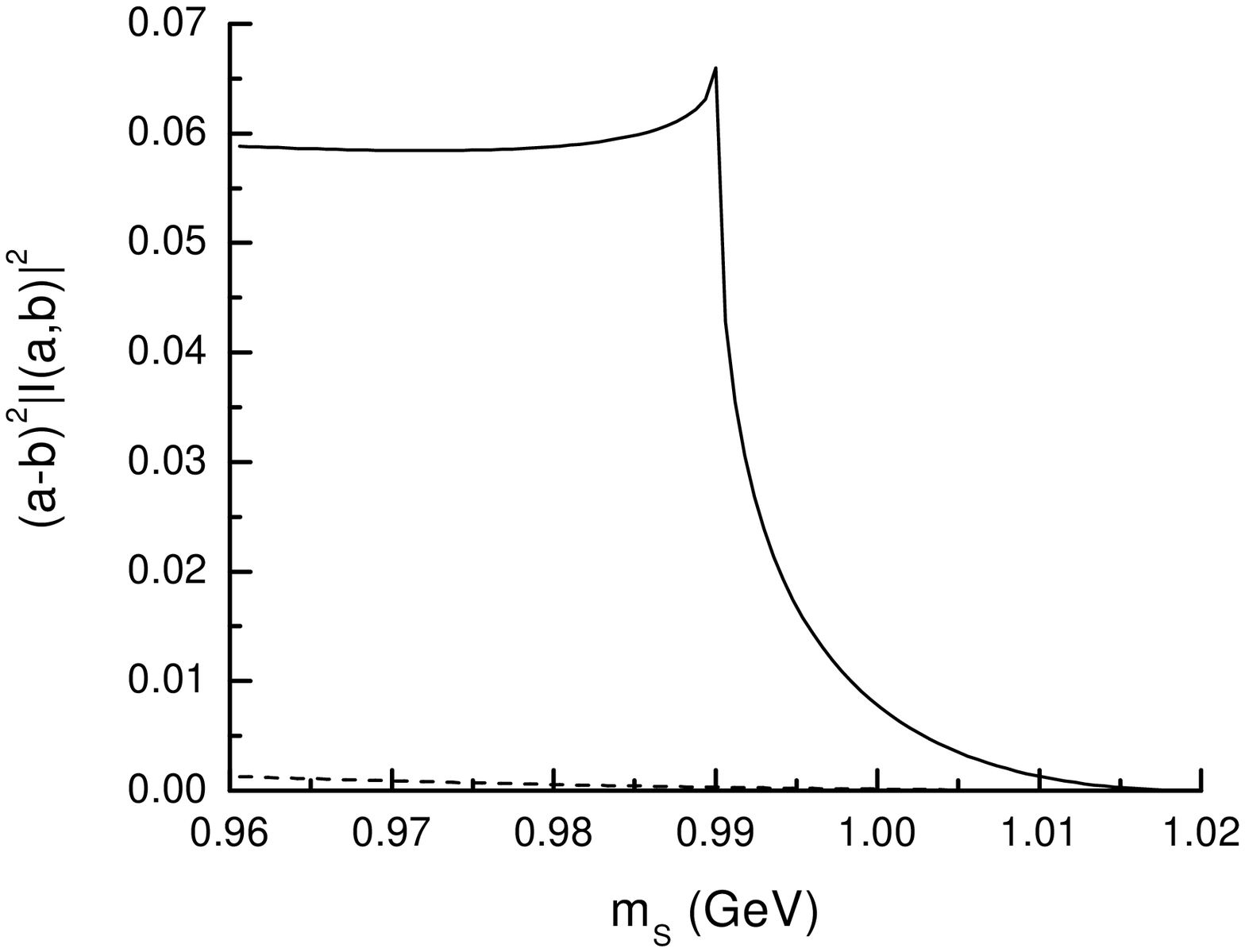,width=6.5cm}\hspace*{2mm}\epsfig{file=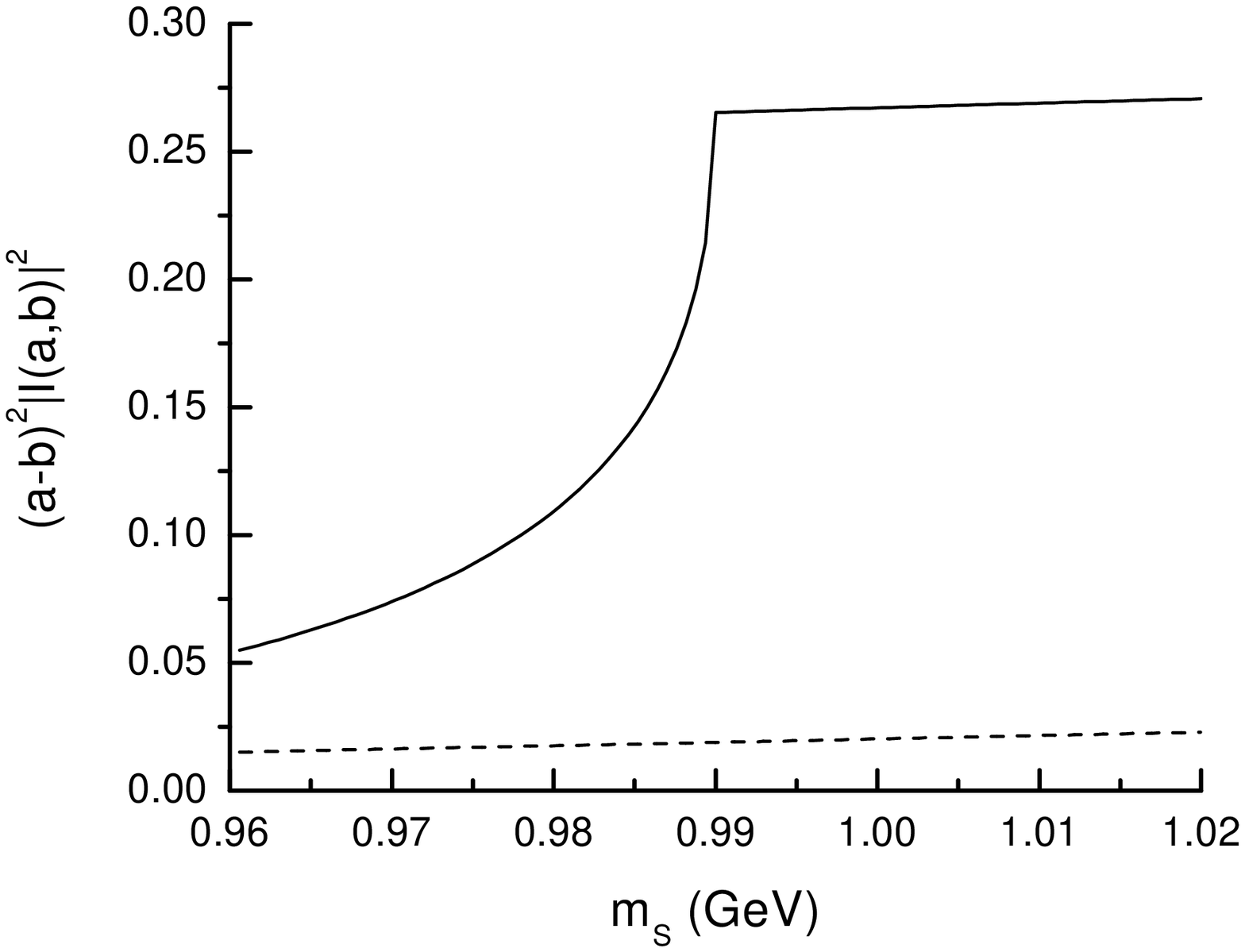,width=6.5cm}} 
\caption{\it Dependence of the function $(a-b)^2|I(a,b)|^2$ on the mass 
of the scalar meson (in GeV) for the kaon loop (solid line) and for the pion 
loop (dashed line). Here $m_V=m_\phi$ 
for the left plot ($\phi$ decay), and $m_V=m_{\rho/\omega}$ for the right plot (scalar decay).} 
\label{s2gamrho}
\end{center}
\end{figure}

The contribution of meson loops is shown diagrammatically in
Fig.~\ref{melo}, where the diagrams $a)$ and $b)$ correspond to the 
coupling of the photon
to the charge of the intermediate pseudoscalar meson and the diagram $c)$ 
stems from gauging the decay vertex of the vector meson to two pseudoscalars. The 
explicit expressions for the corresponding matrix elements read
\begin{eqnarray*}
W_a=W_b=-eg_Sg_V\int\frac{d^4l}{(2\pi)^4}\frac{\epsilon^{\gamma*}{\cdot}(p+q-2l)\;
\epsilon^{V*}{\cdot}(2l+q)}{((p-l)^2-m_P^2)((q-l)^2-m_P^2)(l^2-m_P^2)},\\[2mm]
W_c=-2eg_Sg_V(\epsilon^{\gamma*}{\cdot}\epsilon^{V*})\int\frac{d^4l}{(2\pi)^4}
\frac{1}{((p-l)^2-m_P^2)(l^2-m_P^2)}. 
\end{eqnarray*}
Adding these three we get for the amplitude $M$ introduced in Eq.~(\ref{vertex}):
\begin{equation}
M(m_V^2,m_S^2)=\frac{eg_Sg_V}{2\pi^2m_P^2}I(a,b),
\label{I}
\end{equation}
where $a=\frac{m_V^2}{m_P^2}$, $b=\frac{m_S^2}{m_P^2}$, with $m_P$ being the mass 
of the pseudoscalar; $g_S$ and $g_V$ are the $SP^+P^-$ and $VP^+P^-$ coupling constants, 
while $I(a,b)$ is the loop integral function. An analytical 
expression for this function can be found, {\em  e.g.}, in Refs.~\cite{AI,CIK}. 
The dependence of $(a-b)^2|I(a,b)|^2$ on the mass of the scalar meson is shown 
in Fig.~\ref{s2gamrho}.

What pseudoscalars can be responsible for the transitions under consideration?
The $a_0$ meson is known to couple to $\pi\eta$ and $K \bar K$, while $\rho$,
$\omega$, and $\phi$ do not couple to $\pi\eta$.  Thus, for the $a_0$, only the
kaon loop is relevant. The $f_0$ meson couples to $\pi\pi$ and $K \bar
K$, whereas the only vector meson coupling to $\pi\pi$ is the $\rho$ meson.
Therefore, for the $f_0$, the pion loop could contribute together with the kaon
loop. Nevertheless, the loop integral depends drastically on the relation
between the initial and final meson masses and on the pseudoscalar threshold.
For both cases of the $\phi \to \gamma a_0/f_0$ and $a_0/f_0 \to \gamma
\rho/\omega$ decays the contribution from the pion loop is small (see
Fig.~\ref{s2gamrho}).  Thus, in what follows, only the kaon loop mechanism is
considered.

The only input needed to evaluate the kaon loop contribution to the radiative
decays are the effective coupling
constants $g_S$ and $g_V$. The decay constant for $\phi \to K^+ K^-$
is readily calculated from the $\phi$ width, 
\begin{equation}
\frac{g_{\phi KK}^2}{4\pi}\approx 1.77,
\end{equation}
and the decay constants for
$\rho/\omega \to K^+ K^-$ can be estimated from that for the $\rho \to \pi
\pi$ decay with the help of $SU(3)$ symmetry considerations yielding
\begin{equation}
g_V=g_{\rho K^+ K^-}=g_{\omega K^+ K^-}=\frac{1}{2}g_{\rho \pi \pi}\approx 2.13,
\quad\frac{g_V^2}{4\pi}\approx 0.36.
\label{gro}
\end{equation}

The last missing ingredient is $g_S$. In Ref.~\cite{radmol}\footnote{The value
given in Ref.~\cite{radmol} should be decreased by a factor of 2 
since only the charged kaons contribute to
the loop mechanism of relevance here.} the value of 
\begin{equation}
\frac{g_S^2}{4\pi}=16m_K\sqrt{\varepsilon m_K}=0.6 \ \mbox{GeV}^2
\label{g_S}
\end{equation}
was estimated. To come to this number the mass of 980 MeV for both
scalars was used which corresponds to a binding energy of 
$\varepsilon=10$ MeV. The quoted estimate is
based on assuming a stable molecule formed by a pointlike
interaction in the $K\bar K$ channel and thus should be viewed as qualitative. 
Such a value of $g_S$ lies within the range given by various parametrisations of the
$a_0/f_0$ propagators existing in the literature --- see Tables in 
Refs.~\cite{Flatte,W}.

Within the $q\bar q$ and the 4--quark pictures, relations between the
effective couplings of the $a_0$ and $f_0$ to $K\bar K$ can be derived readily.
In the $q \bar q$ picture, the scalar mesons are $^3P_0$ states and, for the 
flavour--independent strong interaction, one should have, approximately, 
\begin{equation}
g_{f_0KK}=\left\{
\begin{array}{ccl}
g_{a_0KK},&\rm for&f_0(n \bar n)\\
\sqrt{2}g_{a_0KK},&\rm for&f_0(s \bar s),
\end{array}
\label{qqconst}
\right.
\end{equation}
though one should keep in mind that effects like the instanton--induced forces 
or an admixture of the scalar glueball in the wave function of the $f_0$ may destroy 
these equalities. 

In the four--quark model, the scalar decays are superallowed and one has, for
the $a_0$ and $f_0$ with the quark content given by Eq.~(\ref{4q}), 
the relation 
\begin{equation}
g_{f_0KK}=g_{a_0KK},
\label{4qconst}
\end{equation}
which may be distorted by the aforementioned mixing of the 
isoscalar $sn\bar{s}\bar{n}$ with the $\sigma$-like state. 

On the other hand,
estimates for the absolute values of the scalar coupling constants $g_S$ 
involve calculations of strong decays of 
quark--antiquark, four--quark or molecular states, which are very model dependent.
One might consider to rely on experimental data for determining the absolute 
values of $g_S$. However, the coupling constants extracted from data are
afflicted with large uncertainties, {\em i.e.}, they exhibit variations 
up to a factor of $2\div 3$ --- see Tables in Refs.~\cite{Flatte,W}.
This is due to the scaling property of the Flatt{\' e} distributions near the 
$K \bar K$ threshold, as discussed in detail in Ref.~\cite{Flatte}.
In the following we use the value of Eq.~(\ref{g_S}) for the 
scalar coupling. 

With the given values for the couplings
one obtains, in the kaon loop model, the values
\begin{equation}
\Gamma(\phi \to \gamma S)=0.6~{\rm keV},
\label{numphi}
\end{equation}
and
\begin{equation}
\Gamma(a_0/f_0 \to \gamma \rho/\omega) = 3.4~{\rm keV}.
\label{numrho}
\end{equation}

One should keep in mind that the results (\ref{numphi}) and (\ref{numrho}) are
obtained by assuming the scalar vertex to be pointlike. The procedure which
allows one to include the effects of finite--range scalar meson formfactors in a
gauge--invariant way is well known \cite{Markushin,CIK,AGS}. As shown in detail
in \cite{radmol}, these corrections are small in the case of the $\phi \to \gamma
a_0/f_0$ decay. The reason for this is the following: the $\phi$ as well as the
$a_0/f_0$ are close to the $K \bar K$ threshold, and the loop integral is
saturated by nonrelativistic values of the loop momentum, $|\vec{k}| \ll m_K$, where
$m_K$ is the kaon mass.  The range of the scalar formfactor is defined by the
range of the force and, in the absence of pion exchange between kaons in the
scalar sector, the latter is obviously larger than the kaon mass. On the contrary, the mass of the
$\rho/\omega$ is significantly smaller than $2m_K$, and the typical values of
the momentum in the loop integral are not that small. Thus, for the decays 
$S\to\gamma\rho/\omega$, one expects corrections due to the finite range of the 
scalar formfactor, which would reduce the pointlike result.
    
\section{Comment on the $\gamma \gamma$  decays of scalar mesons}
\label{gamgam}

The transition $a_0/f_0 \to \gamma \gamma$, closely related to the 
class of the reactions $S\to\gamma V$, probes the
matrix element $M(p^2,q^2)$ in a kinematical regime quite different from 
the decays discussed above. 

The $\gamma\gamma$ decay can proceed via the quark loop mechanism. Nonrelativistic quark 
model estimates give \cite{qgamgam} 
\begin{equation}
\Gamma_{\gamma \gamma}(^3P_0)=\frac{15}{4}\Gamma_{\gamma \gamma}(^3P_2)=432\alpha^2\langle Q^2\rangle^2
\frac{|R'(0)|^2}{M_0^4},
\label{nqgamgam}
\end{equation}
where $R(r)$ is the radial part of the wave function, and 
$M_0$ is the mass of the $P$-wave quark--antiquark state, which, in the leading nonrelativistic approximation, is supposed to be
the same for all the members of the $P$-wave multiplet.

The calculation of the squared charge factors 
$\langle Q^2\rangle^2$ yields for the isospin ratios
\begin{equation}
\Gamma[(a_0 \to \gamma \gamma):(f_0(n \bar n) \to \gamma \gamma):(f_0(s \bar s) \to \gamma \gamma)]=
9:25:2.
\label{qgamgamratio}
\end{equation}

Therefore, one can try to estimate the decay width for the scalar $f_0(980)$ from 
the width of the tensor $f_2(1270)$, which is known to be a 
good $n \bar n$ state. The PDG \cite{PDG} quotes
\begin{equation}
\Gamma(f_2(1270) \to \gamma \gamma)=2.61 \pm 0.30~{\rm keV},
\label{f2}
\end{equation} 
that gives\footnote{Although the final state contains two photons and, therefore, 
the matrix element scales as $\omega^2$, the phase space brings the factor of 
$\omega/M^2$, with $M=2\omega$ being the physical quarkonium mass
in its rest frame. Therefore, the relation $\Gamma\propto\omega^3$ holds.}
\begin{equation}
\Gamma(f_0(980) \to \gamma \gamma)=\frac{15}{4}\left(\frac{M(f_0)}{M(f_2)}\right)^3
\Gamma(f_2(1270) \to \gamma \gamma)=4.5~{\rm keV}.
\label{qf0}
\end{equation}
Similar results were obtained in other computations of 
$\Gamma(f_0(980) \to \gamma \gamma)$ based on the $q\bar q$ model of the scalar
mesons \cite{Barnes,Narison}.

In the $qq \bar q \bar q$ picture, the predictions appear to be of the order of 
$0.3$ keV for both the $f_0$ and the $a_0$ \cite{ADS}.
 
The $\gamma \gamma$ decay can also proceed via the kaon loop mechanism (see,
{\em e.g.}, Ref. \cite{mexico}), with the matrix element given by Eq.~(\ref{I}) with
$m_V^2=0$. For a pointlike scalar with the mass of $980$ MeV and
$g_S^2/4\pi$ = 0.6  GeV$^2$ one obtains
\begin{equation}
\Gamma(S\to\gamma\gamma)\approx 0.24~{\rm keV}.
\label{loopgamgam}
\end{equation}
In line with the reasoning of the previous section, this value comes out as our
prediction for the $\gamma \gamma$ decay of a molecule.
However, in the kinematical regime of the $S \to \gamma \gamma$ transition, the
momenta in the kaon loop are in the order of the kaon mass and, as in
the case of the $S \to \gamma V$ transitions, one expects corrections due to
the finite range of the form factor at the scalar vertex.
Indeed an explicit calculation within a molecular model of the scalar mesons 
\cite{OllerOs} yields results for the decay widths 
(0.20 keV for the $f_0$ and 0.78 keV for the $a_0$) that differ from
our predictions, but are still in remarkable qualitative agreement 
with them given the simplicity of our approach. 

The result of Eq. (\ref{loopgamgam}) as well as the applied technique is very
different from those in Refs. \cite{Barnes,Krewald}, where also the
$\gamma \gamma$ width of scalar molecules was calculated.
The authors obtain $\Gamma(f_0(K \bar K) \to \gamma \gamma)=0.6~{\rm keV}$ 
\cite{Barnes}
and $6$ keV \cite{Krewald}. In these references, similarly to the 
positronium $\gamma\gamma$ decay, the transition matrix element is taken 
proportional to the value of the $K \bar K$ wave function at the origin.
Not only is this quantity model dependent (as reflected in the order of 
magnitude variation of the calculated widths), but also we suppose that the 
validity of such an approach is highly questionable for the 
considered decays: the range of the $K\bar K\to \gamma \gamma$ transition
operator is of the same order as that of the wave function.

The experimental values for the $\gamma \gamma$ widths of scalars are \cite{PDG} 
\begin{equation}
\Gamma_{\gamma \gamma}(f_0(980))=0.39^{+0.10}_{-0.13}~{\rm keV},\quad
\Gamma_{\gamma \gamma}(a_0(980))=0.30\pm 0.10~{\rm keV}.
\label{expgamgam}
\end{equation}
The estimates based on the nonrelativistic quark loop (\ref{qf0}) are 
in clear disagreement with these data. 
Although relativistic corrections to the 
formula (\ref{nqgamgam}) evaluated in Ref.~\cite{BCL} reduce the ratio 
$\Gamma_{\gamma \gamma}(^3P_0)/\Gamma_{\gamma \gamma}(^3P_2)$ 
by a factor of 2, this result
is still much larger than the experimental values (\ref{expgamgam}). 
On the other hand, the kaon loop mechanism estimate (\ref{loopgamgam}) 
is certainly compatible with them. Moreover, as shown 
recently in Ref.~\cite{achasovgamgam}, the new data \cite{Belle} for the 
reaction $\gamma \gamma \to \pi^+\pi^-$ in the vicinity of the $f_0(980)$ 
resonance can be described with the kaon loop mechanism using the 
weight factor $D_S(m^2_S)$ which reproduces the $S$-wave $\pi\pi$ 
scattering data. 

Concluding, one can say that the existing data on the $\gamma\gamma$ widths of scalars 
seems to favor a molecular structure of the $f_0/a_0$ mesons. However, one has to
admit that, at present, no reliable estimation of the theoretical uncertainty 
involved in the value quoted in Eq.~(\ref{loopgamgam}) can be given. 

\section{Discussion}

\begin{figure}[t]
\begin{center}
\epsfig{file=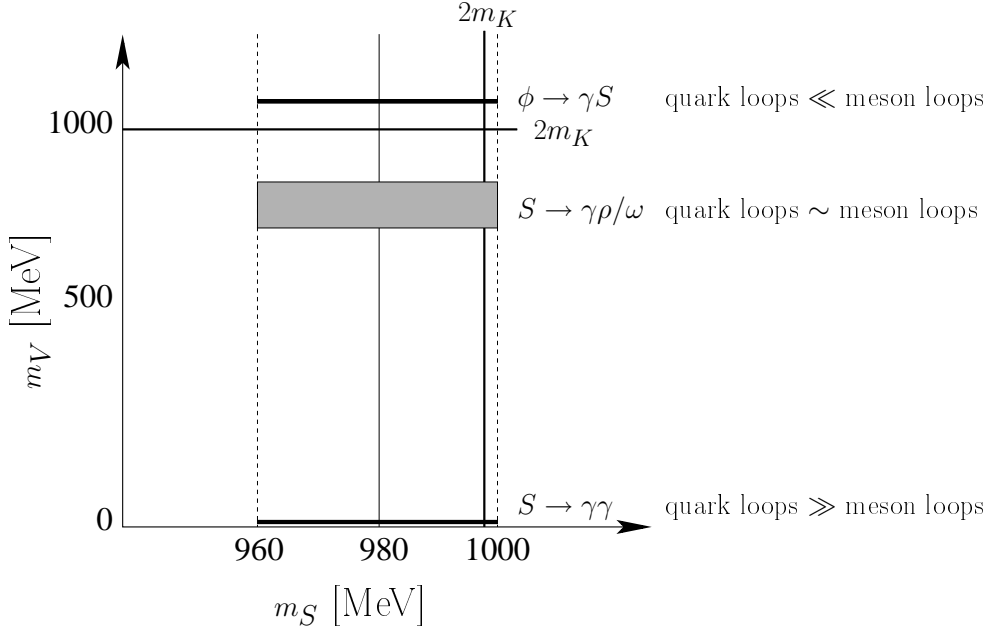, width=13cm}
\caption{\it Illustration of various kinematical regimes probed by the
decays involving scalars.}
\end{center}
\label{melo2}
\end{figure}

The results of the previous sections are summarised in Fig.~\ref{melo2}, where, for the sake of
transparency, the relative contributions of the quark loop and kaon loop mechanisms are 
displayed for various kinematical regimes probed by the radiative decays 
involving scalar mesons.

The message disclosed by this figure is quite clear: the closer the mass of
the vector meson is to the $K \bar K$ threshold, the larger is the
contribution of the kaon loop mechanism to the decay amplitude. Now recall
that the quark loop mechanism is of relevance only if the scalars indeed carry
a significant quark component, whereas the kaon loop mechanism contributes in
both cases, {\em i.e.}, in the $q \bar q$ (or $qq \bar q \bar q$) as well as in the $K
\bar K$ molecule scenario. Thus, in order to discriminate between these two
scenarios, it is most promising to study those decays where the quark loop
mechanism, if present, is significant.

The data on $\phi$ radiative decays yield \cite{PDG}:
\begin{equation}
\begin{array}{l}
Br(\phi \to \gamma a_0) \simeq 7.6\cdot 10^{-5},
\quad\Gamma(\phi \to \gamma a_0)\simeq 0.3~{\rm keV},\\
Br(\phi \to \gamma f_0) \simeq 4.4 \cdot 10^{-4}, 
\quad\Gamma(\phi \to \gamma f_0)\simeq 1.9~{\rm keV},
\end{array}
\label{data}
\end{equation}
indicating that the kaon loop mechanism indeed dominates the radiative
transition $\phi \to \gamma S$.  
This can be understood from the proximity of both the mass of the $\phi$ and
the mass of the scalar mesons to the $K \bar K$ threshold. The question that
remains to be addressed is, however, how much room is there for an additional
quark component. 
We argue that we can get a handle on this quark component of the 
scalar structure when looking at the decays $S \to \gamma \rho/\omega$, for there 
the kinematical situation is quite different. 
As seen from the estimates given by Eqs.~(\ref{V}) and (\ref{numrho}), in
those decays the quark loop and kaon loop
mechanisms should yield contributions of the same order. Accordingly, for
scalar mesons with a $q\bar{q}$ structure, where both
mechanisms contribute, the radiative transition widths should be significantly larger than
for $K \bar K$ molecules, where only the kaon loop mechanism can occur. Thus,
the radiative decays $a_0/f_0 \to \gamma \rho/\omega$ appear to be a much more
decisive testing ground for discriminating between models for the scalar mesons
than the radiative $\phi$ decays.

The estimates of the radiative decay widths of scalar mesons in various models are 
collected in Table~\ref{table1}.
The numbers demonstrate that the $q\bar{q}$ component of 
the scalar mesons implies characteristic ratios for the radiative decays 
into isovector or isoscalar vector mesons, namely 
$\Gamma (a_0 \to \gamma \rho)/\Gamma (a_0 \to \gamma \omega) \approx 1/10$ 
and 
$\Gamma (f_0 \to \gamma \rho)/\Gamma (f_0 \to \gamma \omega) \approx 10$. 
This is in strong contrast to the corresponding ratios for the meson loop
contributions (driven by the kaon loops), where all transitions are
predicted to be of the same order of magnitude so that those ratios should be
in the order of 1. 

The latter point means that the scalar radiative transition is a {\it filtering reaction}. 
The quark loop mechanism ``senses'' the $q \bar q$ flavour, with the ratios of rates for 
different isospin content given by Eq.~(\ref{quarkratio}). 
Thus, one is able to measure the $q \bar q$ content of a specific scalar meson 
produced in a specific reaction
simply by measuring the ratio of decay rates $(f_0 \to \gamma \rho)/(f_0
\to \gamma \omega)$ or $(a_0 \to \gamma \omega)/(a_0 \to \gamma \rho)$.

\begin{table}[t]
\begin{center}
\begin{tabular}{|c|c|c|}
\hline
Decay mechanism&Process&Radiative width in keV\\
\hline
quark loop&$a_0 \to \gamma \rho$&14\\
&$a_0 \to \gamma \omega$&125\\
&$f_0(n \bar n) \to \gamma \rho$&125\\
&$f_0(n \bar n) \to \gamma \omega$&14\\
\hline
$K \bar K$ loop&$a_0/f_0 \to \gamma \rho/\omega$&3\\
\hline
\end{tabular}
\end{center}
\caption{Radiative $S \to \gamma\rho/\omega$ transition in various 
models.}
\label{table1}
\end{table}

Another advantage of the radiative scalar decays is related to the fact that
the phase space available for the final state is not small, in contrast to the
radiative decays of the $\phi$ meson. Thus, simultaneous studies of data on
$\phi$ radiative decays and scalar radiative decays could be useful in
establishing such fundamental characteristics of the scalar mesons as their
pole positions and coupling constants. Indeed, the kaon loop mechanism is
dominant in the $\phi$ radiative decay.  The transition matrix element in the
kaon loop mechanism exhibits a rather peculiar dependence on the masses of the
initial and final mesons. The photon emitted in the $\phi$ radiative decay is
relatively soft, $\omega\sim 40$ MeV, so that the corresponding matrix element
decreases very rapidly in the upper part of the scalar--mass range, from the
$K \bar K$ threshold to the mass of $\phi$ (see the first plot in Fig.~\ref{s2gamrho}). 
On the other hand, in the reaction $S \to \gamma
\omega/\gamma\rho$, the photon energy appears to be large, about $200$ MeV, so that
the matrix element exhibits a rather different pattern.  In the quark loop
model it is nearly constant.  As far as the kaon loop mechanism is concerned,
the corresponding matrix element decreases rapidly with the scalar invariant
mass from the $K \bar K$ threshold downwards, but remains practically constant
above the $K \bar K$ threshold (see the second plot in Fig.~\ref{s2gamrho}). In
such a case it is then possible to analyse the upper part of the spectrum with
much less uncertainty than in the $\phi$ radiative decay.

For the sake of completeness we would like to mention that 
the radiative decays of scalar mesons have been also studied 
within the Vector Dominance Model. Corresponding results can be found, 
e.g., in Ref. \cite{VDM}.

\section{Summary}

We have demonstrated that for radiative decays of the 
scalar mesons $f_0(980)$ and $a_0(980)$ to the vector mesons
$\rho$ and $\omega$, meson loops and quark loops lead to very 
different predictions for the ratio of the decays to $\rho$ and $\omega$,
respectively. Specifically, it follows from our results that 
for objects with a significant component from a compact quark state 
both types of loops should be equally significant. 
On the other hand, for scalar
mesons that are $K\bar K$ molecules only meson loops can contribute. 
The inferred estimates for the decay rates and, in particular, the 
ratios that follow for the two scenarios are so drastically different
that it should be possible to discriminate between them once
experimental information becomes available. 

We have also pointed out that the radiative decay rates involving the scalar mesons 
(such as $\phi\to \gamma S$, $S\to\gamma V$, and $S\to \gamma \gamma$) exhibit a
distinct hierarchy pattern for a compact as well as for a molecular
structure of the scalars. This pattern can be likewise used to distinguish
between the two scenario, and, as an ultimate goal, to define the 
admixture of the bare confined state in the wave function of the scalar
mesons. It requires, however, that a detailed and consistent calculation of all 
those rates is performed within a particular model for the scalar mesons. 
In this context let us mention that the molecular picture of the
scalar mesons has been already successfully tested for the decays $\phi\to
\gamma S$ and $S\to \gamma \gamma$, for which experimental data are available.
Note that there exist calculations \cite{Anisovich} which reproduce both the
value of $2.61$~keV (see Eq.~(\ref{f2})) for the $\gamma\gamma$ width of the 
$f_2(1270)$ and also the $f_0(980)$ data (see Eq.~(\ref{expgamgam})), 
in contrast to results of the
nonrelativistic quark model (\ref{qf0}). Obviously, it would be important to
perform calculations within the quark-model picture of the scalar mesons
for the other decays discussed in this work.

Thus, experimental data on the transitions 
$a_0/f_0\to \gamma \rho/\omega$ --- especially when analysed together 
with the existing data on $\phi\to \gamma a_0/f_0$ --- will
provide strong constraints on models for the structure of the scalar 
mesons and, therefore, are an important source of information 
towards a solution of the scalar-meson puzzle.
\vspace{1cm}

\noindent
{\bf Acknowledgment}

\noindent
The authors would like to thank D. Bugg and E. Oset for a careful
reading of the manuscript and for instructive comments. They also
acknowledge useful discussions with M. B\"usch\-er.
This research was supported by the RFFI via grant 05-02-04012-NNIOa, by the DFG
via grants 436 RUS 113/820/0-1(R), 
by the Federal Programme of the Russian Ministry of Industry, Science, and
Technology No. 40.052.1.1.1112, and by the Russian Governmental Agreement
N 02.434.11.7091. A.E.K. acknowledges also partial support by 
the DFG grant 436 RUS 113/733. 

\appendix

\section{Differential width for unstable particles}

For the calculation of observables, like the decay width, in addition to $M$,
two more ingredients are relevant --- namely, the propagator of the scalar meson
$D_S(p^2)$ and that of the vector meson $D_V(q^2)$. 
The latter modifies the invariant mass spectrum of the final state and 
one can use the unitarity relation to introduce the spectral function $\rho_V$ for the
vector meson,
\begin{equation}
(2\pi)^3\int d\Phi_k(p_V;p_1,\dots,p_k)\, \, |D_V(q^2)W_V|^2=-\frac{1}{\pi}{\rm Im}D_V(q^2)=:\rho_V(q^2),
\end{equation}
where the integral denotes the integration over the phase space of the
decay products of the vector meson that emerged from the vertex $W_V$. 
Note that the spectral density is normalized as
\begin{equation}
\int \rho_V(q^2)dq^2=1.
\label{rhonorm}
\end{equation}
The $\omega$ meson is quite narrow, so that choosing a Breit--Wigner form for
$\rho_\omega$ is appropriate. For the $\rho$ meson
either a Breit--Wigner form or the data from $e^+e^-\to \pi^+\pi^-$ directly can be used.
 
The finite width of the scalars makes one to study the decay rates as a
function of the invariant mass of the decaying system. Consequently, in the total decay width, 
$D_S(m_S^2)$ appears as a weight factor, in the $m^2_S$ integration.
For this distribution of the scalar mesons one would either need to use
parameterizations given in the literature or refer to data from the same
production reaction where the radiative decay is extracted from.

Having this in mind we can straight forwardly generalize Eq. (\ref{observ}) 
to the case of unstable particles in the final and initial state:
\begin{equation}
\frac{d^2\Gamma}{dq^2dm_S^2} =
\frac{m_S^3}{32\pi}|M(m_S^2,q^2)D_S(m_S^2)|^2\left(1-\frac{q^2}{m_S^2}\right)^3\rho_V(q^2) \ .
\label{observ2}
\end{equation}
As mentioned before, the transition matrix element $M$ is the quantity of
interest and we investigate it for various scenarios in the main text.  The
theoretical predictions for the corresponding two dimensional distributions
can be easily generated for each szenario discussed in the main text.

It will be quite demanding to observe the radiative decays of $a_0$ and $f_0$
experimentally. First of all one has to identify reactions that allow one to
disentangle the isoscalar $f_0$ and the isovector $a_0$. Possible reactions
that isolate, e.g., the former state would be $dd\to \alpha+$scalar and $J/\Psi \to
\phi/\omega$+scalar. Then the intermediate scalar states needs to be
reconstructed from the four--vectors of the decay particles.

\end{document}